\documentclass[conference,10pt,letterpaper]{IEEEtran}
\usepackage{amsmath,amssymb,amsfonts,amsthm} 
\usepackage{graphicx,float} 
\usepackage[bookmarks=false]{hyperref} 

\hypersetup{
    hidelinks,              % Removes the colored boxes around links
    draft,                  % A heavy-handed way to strip all link annotations
    breaklinks=true,        % Allows URLs to break over lines without clicking
}
\usepackage{multirow}
\usepackage{tablefootnote}
\usepackage{balance} % This one generates unaccepted IEEE submission format
\usepackage{stfloats}

\usepackage{enumitem, booktabs, textcomp, xcolor, pifont, comment, layout}

\usepackage{caption}
\usepackage{subcaption}

\begin{document}
\title{LLM‑Based Net Analyzer rApp for Explainable and Safe Automation in O‑RAN Non‑RT RIC}

\author{

\IEEEauthorblockN{
Tuan V. Ngo\IEEEauthorrefmark{1}, 
Mao V. Ngo\IEEEauthorrefmark{1},  
%Phuong-Nam Nguyen\IEEEauthorrefmark{1},
Binbin Chen\IEEEauthorrefmark{1}, 
Tony Q. S. Quek\IEEEauthorrefmark{1},
Tejaswita Kumari\IEEEauthorrefmark{2},
and Maziar Nekovee\IEEEauthorrefmark{2}
}
\IEEEauthorblockA{\IEEEauthorrefmark{1}%
Singapore University of Technology and Design (SUTD), 487372, Singapore\\ 
{\{vantuan\_ngo, vanmao\_ngo, 
%nam\_nguyen, 
binbin\_chen, tonyquek\}@sutd.edu.sg}
}

\IEEEauthorblockA{\IEEEauthorrefmark{2}%
                    6G Lab, University of Sussex, UK
{\{tk468, m.nekovee\}@sussex.ac.uk}
}
}

\maketitle

\begin{abstract}
Modern 5G/6G radio access networks are increasingly programmable through O-RAN, yet their operational complexity has grown with disaggregation, open interfaces, and fine-grained control parameters. While RAN-side analytics and telemetry mechanisms, such as KPI-based monitoring and mobility event reporting, provide visibility into network behavior, operators still face challenges in correlating heterogeneous events and safely translating observations into actionable configuration changes. This paper presents an LLM-based Net Analyzer rApp for the O-RAN Non-RT RIC that enables explainable and safe, human-in-the-loop automation for RAN operations. The proposed rApp adopts an event-informed, batch-triggered reasoning framework in which mobility events are first interpreted, anomalies are confirmed through targeted log inspection, configurations are inspected via tool-gated access, and minimal configuration changes are proposed only after explicit operator approval. The architecture enforces a strict separation between reasoning and actuation, ensuring auditability and operational safety. The system is implemented and demonstrated on a real O-RAN testbed using a reproducible ping-pong handover scenario, illustrating how large language models can function as reasoning co-pilots that transform raw RAN telemetry into structured explanations and controlled remediation workflows, complementing existing analytics-only approaches in the Non-RT RIC.
\end{abstract}


\section{Introduction}

Modern 5G and emerging 6G Radio Access Networks (RANs) are increasingly software-defined and data-driven. 
At the same time, their operational complexity has grown due to RAN disaggregation, open interfaces, and fine-grained configuration parameters. 
The Open RAN (O-RAN) architecture exposes programmability through the Non-Real-Time RAN Intelligent Controller (Non-RT RIC), which hosts rApps, the Near-Real-Time RAN Intelligent Controller (Near-RT RIC), which hosts xApps, and standardized interfaces such as O1, A1, and E2. 
While this enables advanced optimization, it also overwhelms operators with heterogeneous events, logs, and measurements that are difficult to correlate and safely translate into control actions. 

In the 5G Core, analytics-driven management is formalized through the Network Data Analytics Function (NWDAF) defined by 3GPP in TS~23.288 \cite{3gpp-ts23288}. 
%NWDAF provides statistical and predictive analytics to network functions and OAM consumers via subscription and query services. 
NWDAF provides statistical and predictive analytics to network functions and OAM consumers but leaves control decisions to external consumers, meaning it does not directly close the loop on configuration updates.
In contrast, we propose an rApp that automatically analyzes RAN events, logs, performance metrics, and incorporates explanation-centric reasoning with tool-gated configuration workflows under human approval. This design moves the O-RAN-based system toward a safer, semi-automated closed-loop operation.
%In contrast, the proposed rApp integrates explanation-centric reasoning with tool-gated configuration workflows under human approval, bringing the system closer to safe, semi-automated closed-loop operation.

Recent work highlights the potential of Large Language Models (LLMs) to assist network operations by interpreting operator intents, correlating heterogeneous telemetry, and orchestrating tool-based diagnosis and configuration synthesis. 
However, prior studies also emphasize risks such as hallucination, unsafe actuation, and lack of auditability, motivating the need for strict safety guardrails and human-in-the-loop control. 
Existing LLM-based approaches rarely demonstrate tool-bounded reasoning integrated with the O-RAN Non-RT RIC or formalize explicit termination and operator approval checkpoints required for production-grade operations. 

This paper addresses these gaps by introducing an \textit{LLM-based Net Analyzer rApp} for the O-RAN Non-RT RIC. 
The proposed rApp operationalizes an LLM as a loop-driven reasoning agent that interprets mobility events, correlates multi-source logs, inspects configurations via tool-gated access, and proposes minimal configuration changes subject to explicit operator approval. 
The design is validated on a real O-RAN hardware 5G testbed using a reproducible ping-pong handover scenario, demonstrating explainable diagnosis and safe, human-in-the-loop remediation. Our contributions in the proposed rApp are as follows: 

\begin{itemize}
    \item \textit{Loop‑driven reasoning with Modes:} We codify a strict EVENT $\xrightarrow{}$ NEXT/HUMAN  loop that first classifies behavior and plans evidence collection, then confirms anomalies from logs, then inspects configuration, and only after explicit operator approval applies minimal changes—terminating with a verified final state. This addresses safety, transparency, and auditability gaps identified in emerging LLM‑Ops workflows for telecom.

    \item \textit{Tool‑gated Access:} The LLM never interacts with live systems directly; it invokes log-query and configuration-query tools with strictly scoped queries and paths, using the O1 interface in accordance with O-RAN practices and avoiding any free‑form actuation.

    \item \textit{Experimental validation on an O-RAN 5G testbed:} the proposed rApp is tested on an O‑RAN 5G testbed under a standards‑compliant mobility scenario. We inject A3 mobility misconfiguration and show the rApp’s capability to detect ping‑pong handovers issue, explain root causes, and prepare minimal configuration patches, and request human operator approval before safely applying updates to the live 5G system via O1 interface. %contrasting with analytics‑only solutions such as NWDAF or KPM‑centric xApps. 
    Experimental results demonstrate clear benefits for both network operators and end-user experience.

    %\item Relative to prior RIC prototypes (e.g., FlexRIC/xApps), our focus is not on adding new service models or Near‑RT control loops, but on human‑explainable diagnosis and safe configuration closure at Non‑RT time scales, complementing Near‑RT controllers.
\end{itemize}

\section{Related Work}
\label{sec:relatedWork}
Prior work on O‑RAN programmability has established the feasibility of deploying xApps and rApps on platforms such as the O‑RAN Software Community (OSC) RIC~\cite{oran_sc_nonrtric_docs} and FlexRIC. Mao \textit{et al.}~\cite{Mao_ICT2024} demonstrate real‑world deployment of such platforms on commercial 5G networks, but these studies primarily validate architecture and service models rather than automated diagnosis or safe configuration workflows. Meanwhile, AI‑driven rApps have been explored for closed‑loop optimization: Long \textit{et al.}~\cite{Long_Infocom_Wkshp2024} deploy task‑specific prediction models for localization and interference management, and Rahman \textit{et al.}~\cite{Rahman_GlobcomWkshp2024} enable slice QoS control using LSTM‑based predictors. These systems rely on black‑box models and KPI‑centric actuation, lacking mechanisms to correlate heterogeneous evidence or provide operator‑interpretable reasoning.

LLMs have recently been introduced into telecom operations for analytics and operator assistance. AutoRAN~\cite{autoRAN} introduced an LLM-assisted, zero-touch provisioning framework in which  high-level intents are translated into machine-readable configuration files to automate multi-vendor O-RAN deployment and orchestration pipelines.
Mobile‑LLaMA~\cite{Kan_MobileLLaMA} integrates an instruction‑tuned LLM with an NWDAF‑aligned pipeline to translate operator queries into executable analysis code, but it targets 5G Core analytics and does not operate on RAN telemetry or integrate with the O‑RAN RIC control framework. Broader architectural perspectives, such as Mekrache \textit{et al.}~\cite{Mekrache_IntentBasedManagement} and LLM‑hRIC by Bao \textit{et al.}~\cite{bao2025llm}, explore LLM‑enabled intent management or hierarchical control through Non-RT RIC and Near-RT RIC. However, these systems emphasize high‑level guidance or simulated performance optimization rather than event‑driven diagnosis, tool‑restricted evidence gathering, or operator‑approved corrective actuation in live O-RAN environments.

Overall, existing solutions either focus on prediction‑driven closed loops, operate outside the RAN telemetry environment, or emphasize architectural workflows without concrete, safe configuration mechanisms. In contrast, our work introduces an LLM‑based rApp integrated into the O‑RAN Non‑RT RIC, equipped with structured, evidence‑grounded reasoning, tool‑gated access to logs and configuration, and explicit human‑in‑the‑loop approval—filling key gaps in explainability, safety, and auditability required for practical RAN operations.

\section{LLM-Based Net Analyzer rApp Architecture and Design}
\label{sec:architecture}

\begin{figure}[t]
  \centering
  \includegraphics[width=0.95\linewidth]{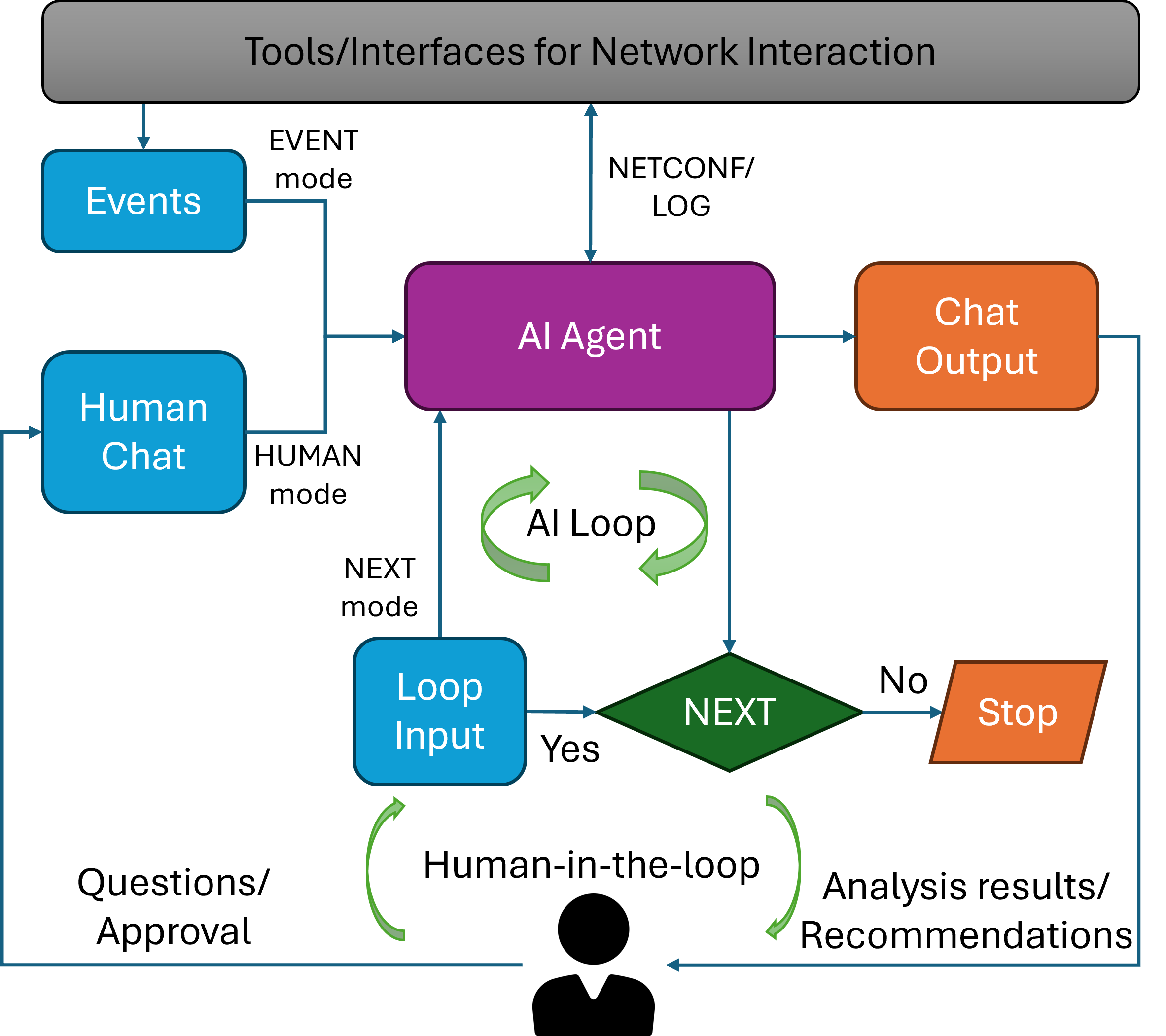}
\caption
{
Simplified architecture of the LLM-based Net Analyzer rApp. 
RAN event inputs and human chat inputs are processed by an LLM-based reasoning agent operating within an explicit event-informed, batch-triggered loop.
%A \texttt{NEXT} transition is taken only when the agent signals continuation and orchestration constraints permit another iteration, while termination occurs explicitly otherwise, ensuring bounded and safe execution.
}
  \label{fig:llm_net_analyzer_arch}
\end{figure}

This section presents the architecture and design principles of the proposed LLM-based Net Analyzer rApp deployed within the O-RAN Non-RT RIC.
The rApp is designed to support explainable, safe, and human-in-the-loop RAN analysis at non-real-time timescales.
Rather than performing autonomous optimization, the rApp operates as a reasoning co-pilot that assists operators in diagnosing network behavior and preparing controlled configuration actions.

\subsection{System Architecture}

Fig. \ref{fig:llm_net_analyzer_arch} illustrates a simplified logical view of the proposed architecture.
The rApp follows an event-informed, batch-triggered design in which RAN telemetry and human inputs are processed by an LLM-based reasoning core.
The architecture separates input handling, reasoning, and execution control to ensure bounded and observable operation.

Two primary input channels are supported: Events, and Human Chat, as shown in blue blocks in Fig.~\ref{fig:llm_net_analyzer_arch}. 
First, Event inputs consist of asynchronous RAN-side telemetry, including events and performance indicators exported through Non-RT RIC. 
However, rather than reacting directly to each raw streaming message, the rApp operates on bounded, normalized batches of events delivered via a queuing and orchestration layer. 
This layer regulates event flow and prevents the LLM from being overwhelmed by high-rate or bursty telemetry.
A reasoning cycle is triggered only when a batching condition is satisfied, such as the absence of new events within a predefined time window or the accumulation of a sufficient number of events in the queue, ensuring bounded and stable reasoning inputs.
Second, Human Chat inputs allow operators to request explanations, provide contextual information, and explicitly approve or reject proposed actions. 

At the core of the rApp is an LLM-based Reasoning Agent, shown as a pink block in Fig.~\ref{fig:llm_net_analyzer_arch}. %instantiated at the Non-RT timescale.
In our implementation, GPT-4.1-mini is used to balance reasoning capability, latency, and operational cost.
The LLM is used exclusively for semantic interpretation, hypothesis formulation, and orchestration of analysis steps---not for direct actuation.

\subsection{LLM-Based Reasoning Framework}

The rApp adopts a structured reasoning framework that constrains the behavior of the LLM. 
%The framework defines explicit execution modes and restricts how the reasoning process progresses.
Three execution modes (refer to Fig.~\ref{fig:llm_net_analyzer_arch}) provide structure and traceability:
%As illustrated in Fig. \ref{fig:llm_net_analyzer_arch}, three execution modes are supported.
\begin{itemize}
    \item EVENT mode: Agent interprets aggregated RAN context and assesses whether the observed behavior deviates from expected operation.
    \item NEXT mode: Agent requests additional evidence or context, such as log inspection or configuration queries, and initiates another iteration of reasoning.
    \item HUMAN mode: Agent presents explanations and proposed actions to the operator and awaits explicit approval or rejection.
\end{itemize}

The LLM is strictly prohibited from directly modifying network configuration or performing irreversible actions.
As shown in Fig.~\ref{fig:llm_net_analyzer_arch}, all interactions occur through tool-gated interfaces consistent with O-RAN management practices, ensuring that configuration updates are executed solely by the orchestration layer following explicit operator approval.
This layer exposes only scoped queries and controlled access paths, preventing the LLM from engaging in unrestricted system interaction. 
As a result, every reasoning step remains observable, auditable, and fully enforceable. % under operator and orchestration control.

\subsection{Event-Informed, Batch-Triggered Automation Loop}

The operation of the Net Analyzer rApp is governed by an explicit event-informed, batch-triggered, loop-driven automation process.
A reasoning cycle begins when new event inputs are received from the RAN.
If the observed behavior is classified as normal, the agent reports the assessment and terminates the cycle.

When abnormal behavior is suspected, the agent formulates an initial hypothesis and provides a preliminary explanation to the operator.
If additional evidence is required, the agent issues a NEXT decision together with a well-defined follow-up action, triggering another iteration with enriched context.
This process continues until sufficient evidence is obtained or operator input is required.

The rApp enforces bounded and auditable execution through a two-layer termination mechanism. 
At the agent layer, the LLM emits a control intent indicating whether further evidence collection or interaction is required. 
At the orchestration layer, this intent is evaluated together with a predefined iteration bound, and a \texttt{NEXT} transition is taken only if both the agent signals continuation and the current iteration count remains below the cap. 
Otherwise, the reasoning loop terminates explicitly. 
This design preserves the semantic notion of an explicit stop while guaranteeing a deterministic upper bound on runtime and tool usage. 
In our implementation, the iteration cap is set to a small constant (e.g., five iterations) and can be adjusted by the operator according to operational policy. 

% Explicit termination ensures bounded execution and predictable system behavior.
% Through this design, the LLM functions as an explainable reasoning co-pilot that supports safe and human-governed RAN operations.

\section{Experimental Setup and Evaluation}
\label{sec:evaluation}

\begin{figure*}[t]
  \centering
  \includegraphics[width=0.970\textwidth]
  %{imgs/oran_testbed_topology.png}
  {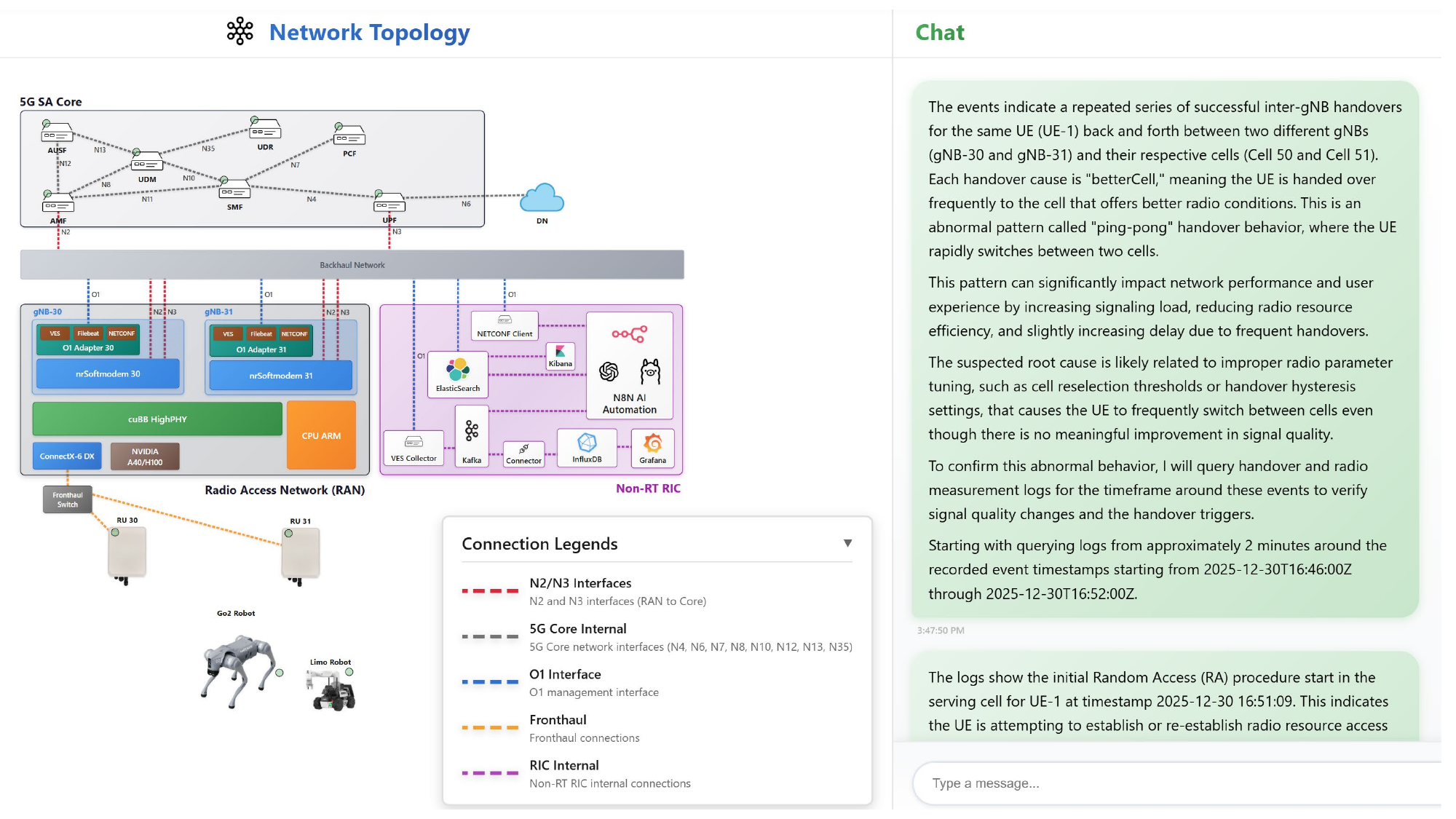}
  \caption{O-RAN testbed and network topology used for evaluating the proposed Net Analyzer rApp.
Two gNBs (gNB-30, gNB-31), each connected to a corresponding RU, form two neighboring indoor cells with partial coverage overlap.
The Non-RT RIC hosts multiple components, including events, log and configuration collectors, data storage backends, and the proposed rApp.}
  \label{fig:oran_testbed}
\end{figure*}

This section describes the experimental environment and evaluation scenario used to demonstrate the feasibility and behavior of the proposed LLM-Based Net Analyzer rApp. 

\subsection{O-RAN Testbed Description}
\label{sec:testbed}

Built upon our AI-RAN testbed~\cite{nguyen2025adaptiveaimodelpartitioning}, Fig.~\ref{fig:oran_testbed} depicts our O-RAN testbed used in the experimental evaluation. The figure shows two gNBs connected to a common 5G Core (5GC) and managed by an O-RAN Service Management and Orchestration (SMO) framework hosting the Non-RT RIC.

On the left side of the figure, two gNB instances are shown, each serving a single 5G NR cell. Both gNBs are implemented using OpenAirInterface (OAI) and share a single CUDA-accelerated High-PHY instance based on NVIDIA Aerial, which executes the baseband processing on NVIDIA GPU GH200. The two gNBs operate as independent logical entities while leveraging the same High-PHY processing pipeline, enabling a controlled multi-cell deployment with a single accelerated physical layer.

Each gNB is connected to a Sera Networks O-RU~\footnote{https://www.sera-network.com/?portfolio=oru-nw4400} through an O-RAN 7.2x fronthaul interface. Radio measurements and mobility-related events are generated at the gNBs as the user equipment (UE) moves between the coverage areas of the two cells. Inter-cell mobility is realized through NG-based handover procedures coordinated by the 5GC, with the source gNB reporting measurement results and handover events, and handover execution and completion managed via the core network.

RAN management and telemetry exposure are handled through an O1 adapter integrated with the gNBs, as illustrated in the figure. The O1 adapter exports mobility events, handover outcomes, and operational logs to the Non-RT RIC, enabling reasoning and analysis by LLM-based Net Analyzer rApp.

The 5GC, shown at the top of the figure, provides the control-plane functions required for NG handover, including UE context transfer and mobility anchoring. While the 5GC participates in the handover signaling path, it does not perform analytics, reasoning, or parameter optimization in this experiment. Its role is limited to enabling standard-compliant mobility execution, ensuring that the observed handover behavior reflects realistic deployment conditions.

On the right side of the figure, based on the OSC Non-RT RIC framework~\cite{oran_sc_nonrtric_docs}, we implemented additional features to support data ingestion, storage components, and include the N8N AI automation framework~\cite{n8n} to implement the LLM-based Reasoning Agent. 
RAN-side events and measurements exported from the gNBs via the O1 adapter are first received by a VES collector, which normalizes mobility events and handover notifications before forwarding them into a Kafka message bus. Kafka acts as the central event queue, decoupling event ingestion from downstream processing and enabling asynchronous consumption by analytics and reasoning components.

Telemetry streams are persisted through dedicated connectors. Time-series measurements are written to InfluxDB via an InfluxDB connector, while structured logs and event records are indexed into Elasticsearch. Kibana, shown alongside ElasticSearch in Fig.~\ref{fig:oran_testbed}, is used for log inspection and visualization during analysis and debugging. These data stores provide historical context that complements real-time event streams.

Implemented using a native AI automation platform (N8N)~\cite{n8n}, the proposed LLM-based Net Analyzer rApp is deployed within the Non-RT RIC and operates on events ingested through O1 interfaces. 
Kafka is used as an upstream message bus to decouple RAN-side event ingestion from downstream processing, while the rApp is triggered by normalized, batched event inputs from the orchestration layer.
In addition to event inputs, the rApp queries ElasticSearch and InfluxDB for relevant logs and measurements to enrich its reasoning process. 
Configuration access paths are also illustrated in the figure, where the rApp interacts with gNB configuration exclusively through NETCONF-based management interfaces exposed by the O1 adapter. 
Rather than directly modifying configurations, the rApp issues configuration read and proposal requests through this path, preserving a clear separation between telemetry ingestion, reasoning, and configuration execution. 

\subsection{Ping-Pong Handover Experiment Setup}
\label{sec:scenario}

\begin{figure}[t]
    \centering
    \includegraphics[width=0.95\linewidth]{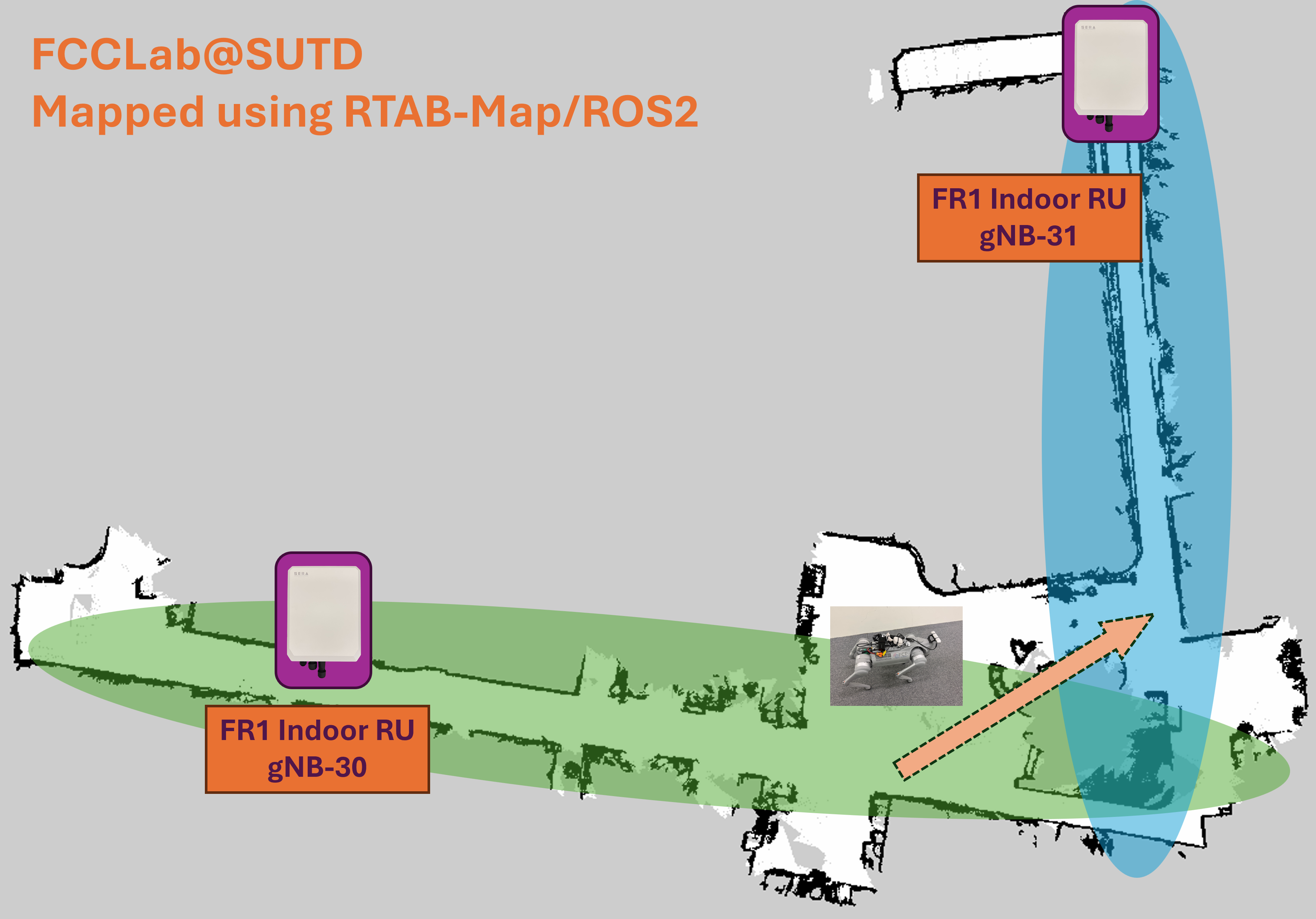}
    \caption{Ping-pong handover experimental scenario: %Two 5G indoor cells served by gNB-30 and gNB-31 provide overlapping coverage. 
    A 5G robot-dog follows the indicated trajectory, repeatedly crossing the cell boundary and triggering successive handovers between the two cells.}
    \label{fig:pingpong_scenario}
\end{figure}

Fig.~\ref{fig:pingpong_scenario} illustrates the ping-pong handover scenario used in the experimental evaluation. Two FR1 indoor radio units are deployed at opposite ends of the environment and are labeled as gNB-30 and gNB-31. 
Each gNB serves a single NR cell, with the corresponding coverage areas shown by the shaded regions in the figure.
Both cells operate on the N78 band with the same 50 MHz bandwidth, resulting in an intra-frequency handover whenever a mobile user equipment (UE) transitions between the two cells. 

The mobile UE---represented by the 5G-enabled robot-dog G02---moves along the trajectory indicated by the arrow, repeatedly traversing the overlapping region between the two cells. 
As the UE approaches the boundary of the serving cell, radio measurements (e.g., RSRP/RSRQ) fluctuate due to comparable signal conditions from both gNBs. 
When handover parameters are not properly configured for both cells, these fluctuations cause frequent triggering of mobility events, resulting in \textit{ping‑pong handovers} as the UE alternates between gNB‑30 and gNB‑31.
%these fluctuations lead to frequent triggering of handover conditions as the UE alternates between the coverage areas of gNB-30 and gNB-31.

During this movement, the 5G UE undergoes rapid and successive handovers between the two cells, producing a characteristic \textit{ping‑pong mobility pattern} within a short time interval. These repeated transitions are clearly reflected in the sequence of mobility events and handover outcome notifications generated by both gNBs and exported to the Non‑RT RIC. The controlled deployment geometry and deterministic UE trajectory, as depicted in Fig.~\ref{fig:pingpong_scenario}, create a highly reproducible test condition that enables rigorous evaluation of the proposed LLM-based Net Analyzer rApp’s ability to \textbf{detect, interpret,} and \textbf{reason} about unstable mobility behavior.

\textbf{Precondition:} Prior to the experiment, we intentionally misconfigure the A3 handover parameters of the two cells (i.e., offset, hysteresis, or time-to-trigger) to induce suboptimal mobility handover. 
We then conduct two runs: a baseline trial without the proposed rApp enabled, followed by a second trial with the rApp active. 
In both cases, we observe the quality of the live video stream from robot-dog camera over 5G network as it moves along a predefined path from one corner to the other at the FCCLab at the Singapore University of Technology and Design, as shown in Fig.~\ref{fig:pingpong_scenario}.  

\textbf{Metrics:} To benchmark the two scenarios, we record the video streaming performance, focusing specifically on the achieved frame‑per‑second (FPS) rate during the robot’s movement.

%During this movement, the 5G UE experiences successive handovers between the two cells within a short time interval, resulting in a ping-pong handover pattern. The repeated transitions are observable in the mobility events and handover outcome notifications generated by the gNBs and exported to the Non-RT RIC. This controlled physical layout and UE trajectory, as shown in Fig.~\ref{fig:pingpong_scenario}, provide a reproducible scenario for evaluating the ability of the proposed Net Analyzer rApp to detect, explain, and reason about unstable mobility behavior.

\subsection{End-to-End rApp Execution}
\label{sec:execution}

\begin{figure}[ht]
\centering
\footnotesize
\setlength{\fboxsep}{5pt}
\setlength{\fboxrule}{0.4pt}

% rApp message 1: observation (not explanation)
\fbox{
\begin{minipage}[t]{0.85\linewidth}
\textbf{Net Analyzer rApp}
\emph{Abnormal mobility detected.}\\
Repeated inter-gNB handovers observed for UE~\#17 within a short time window.
This pattern is inconsistent with stable mobility behavior.\\
\textbf{NEXT:} Inspect handover logs.
\end{minipage}
}

\vspace{0.5em}

% rApp message 2: causal evidence
\fbox{
\begin{minipage}[t]{0.85\linewidth}
\textbf{Net Analyzer rApp}
\emph{Log inspection completed.}\\
Handover logs show repeated Event~A3 triggers caused by marginal RSRP differences ($<1$~dB).
This condition is indicative of ping-pong handover behavior.\\
\textbf{NEXT:} Inspect Event~A3 configuration.
\end{minipage}
}

\vspace{0.5em}

% rApp message 3: root cause
\fbox{
\begin{minipage}[t]{0.85\linewidth}
\textbf{Net Analyzer rApp}
\emph{Configuration inspection completed.}\\
Current Event~A3 offset, hysteresis, and time-to-trigger values are overly sensitive.
This sensitivity increases susceptibility to ping-pong handovers under marginal signal fluctuations.\\
\textbf{HUMAN:} Recommend updated Event~A3 parameters.
\end{minipage}
}

\vspace{0.5em}

% Human asks for recommendation
\hfill
\fbox{
\begin{minipage}[t]{0.6\linewidth}
\textbf{Human Operator}\\
What configuration values do you recommend?
\end{minipage}
}

\vspace{0.5em}

% rApp recommendation
\fbox{
\begin{minipage}[t]{0.85\linewidth}
\textbf{Net Analyzer rApp}
\emph{Recommended values.}\\
Increase Event~A3 offset from 2$\rightarrow$4~dB.\\
Increase hysteresis from 2$\rightarrow$4~dB.\\
Increase time-to-trigger from 100$\rightarrow$320~ms
to reduce sensitivity to transient signal variations.\\
\textbf{HUMAN:} Await human approval.
\end{minipage}
}

\vspace{0.5em}

% Human approval
\hfill
\fbox{
\begin{minipage}[t]{0.6\linewidth}
\textbf{Human Operator}\\
Approve.
\end{minipage}
}

\vspace{0.5em}

% rApp final message
\fbox{
\begin{minipage}[t]{0.85\linewidth}
\textbf{Net Analyzer rApp} \emph{Configuration update confirmed.}\\
Event~A3 parameters have been updated by the orchestration layer following operator approval.\\
Observed mobility behavior is expected to stabilize.\\
\textbf{STOP:} Terminate reasoning cycle.
\end{minipage}
}

\caption{Simplified dialogue between Net Analyzer rApp and Human Operator during resolving the ping-pong handover issue.}
\label{fig:chat_exchange}
\end{figure}

Fig.~\ref{fig:chat_exchange} shows a simplified chat-style interaction between the LLM-based \textit{Net Analyzer rApp} and a \textit{Human Operator} during the resolution of the end-to-end ping-pong handover issue.
The conversation captures a complete reasoning cycle triggered by mobility events, including evidence collection, configuration sensitivity analysis, and the generation of concrete parameter recommendations.
In this work, an explanation refers to a causal reasoning trace that links observed RAN events to specific triggering conditions and configuration sensitivities, thereby justifying a concrete, human-approved configuration action.
Each rApp message explicitly specifies the \emph{NEXT} decision, making the progression and termination of the reasoning loop transparent and auditable.
Human involvement is limited to clarification, recommendation requests, and final approval, while intermediate reasoning steps are performed internally by the LLM through its controlled interaction with the O-RAN system.

\subsection{Experimental Results and Discussion}
\label{sec:observations}

\begin{figure}[ht]
    \centering
    \includegraphics[width=0.95\linewidth]{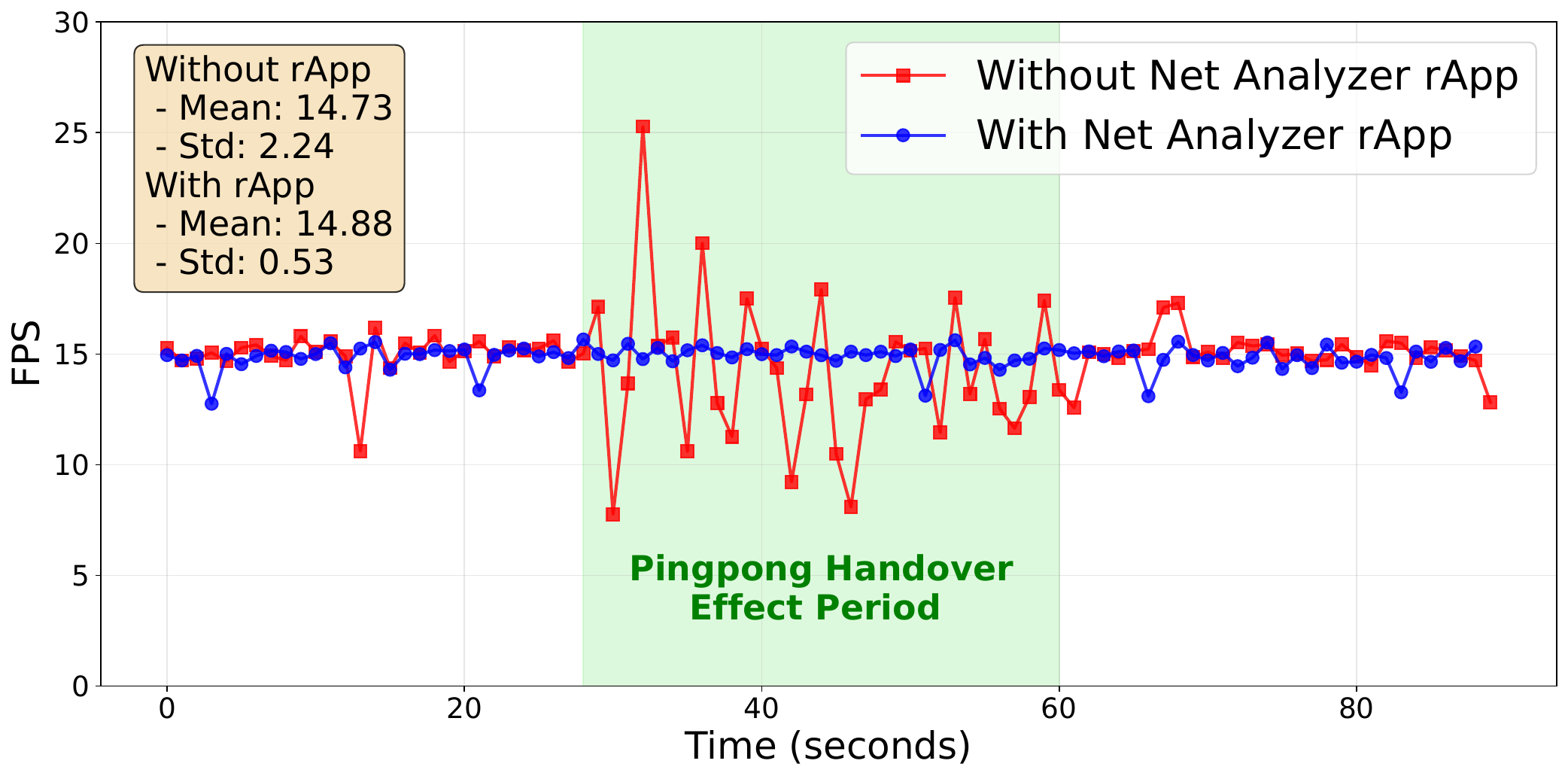}
    \caption{Video streaming frame per second (FPS) comparison between the two scenarios with/without the proposed rApp enabled.}
    \label{fig:fps_comparison}
\end{figure}

Fig.~\ref{fig:fps_comparison} shows the video‑streaming frame rate (FPS) measured at the remote controller as the 5G robot-dog transmits uplink video during teleoperation, both before and after applying the configuration adjustments recommended by the LLM‑based \textit{Net Analyzer rApp} in the ping‑pong handover scenario. In this experimental setup, the UE‑mounted camera continuously sends real‑time video over the O‑RAN‑based 5G testbed, and the remote controller relies on this stream as its primary source of visual feedback for steering and task execution.

%Fig.~\ref{fig:fps_comparison} presents the video streaming frame rate (FPS) observed at the remote controller for a video stream transmitted from the 5G robot dog during remote robot operation, before and after applying the configuration changes recommended by the LLM-based Net Analyzer rApp in the ping-pong handover scenario.
%In this setup, the UE-mounted camera continuously streams video over the O-RAN-based 5G testbed to a remote controller, which uses the stream as the primary visual feedback for teleoperation.

Under the scenario without the proposed rApp, frequent ping‑pong handovers cause recurring uplink interruptions as the UE moves across the boundary between gNB‑30 and gNB‑31. These interruptions manifest as noticeable fluctuations in FPS at the remote controller---highlighted by the red line in Fig.~\ref{fig:fps_comparison} between seconds 25 and 60---disrupting visual continuity and increasing operator uncertainty during closed‑loop robot motion. As a result, instability at the mobility layer directly propagates to the application layer, degrading the overall user‑perceived quality of teleoperation.

%As shown in Fig. \ref{fig:fps_comparison}, under the baseline configuration, frequent ping-pong handovers introduce repeated uplink disruptions during UE mobility. These disruptions manifest as pronounced FPS fluctuations at the remote controller, degrading visual continuity and increasing the operator's control uncertainty.

\begin{figure}[ht]
    \centering
    \includegraphics[width=1.0\linewidth]{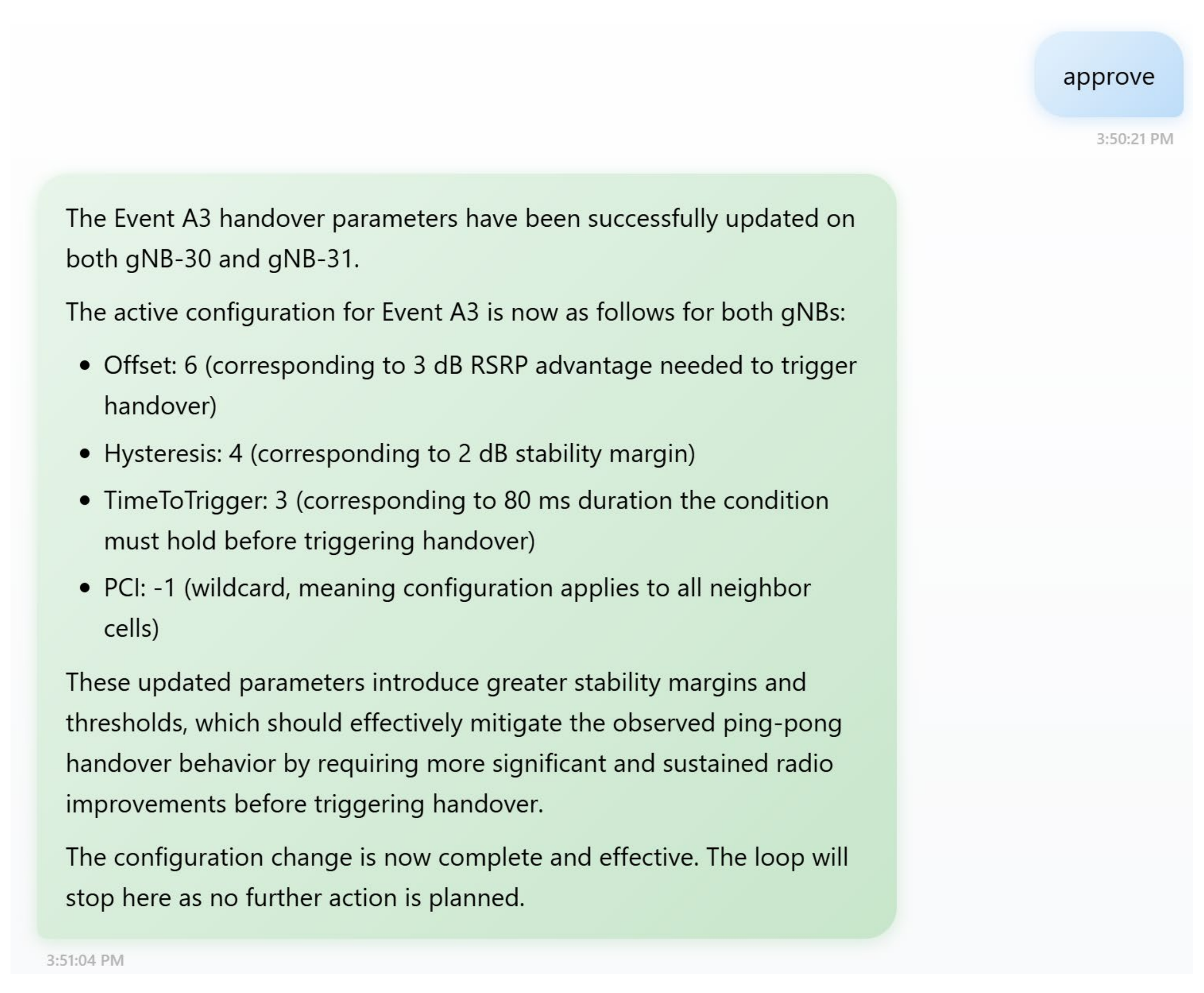}
    \caption{Example chatbot in LLM-based Net Analyzer rApp interaction showing human approval followed by the successful update of corrected configurations to the two gNBs.}
    \label{fig:output_after_approve}
\end{figure}

Under the scenario with the proposed rApp, we can observe FPS performance shown by the blue line in Fig.~\ref{fig:fps_comparison}. 
After the rApp analyzes abnormal mobility events and proposes adjustments to handover parameters, the operator-approved configuration changes are applied to the network.
As shown in Fig.~\ref{fig:output_after_approve}, after operator \textit{approve}s, the LLM-based rApp updates the handover A3 Event parameters (i.e., Offset=6, Hysteresis=4, Time-to-trigger=3) to both gNB-30 and gNB-31 via O1 interface.
Following this intervention, the uplink video stream exhibits a markedly more stable FPS profile, with reduced variance and fewer short-term degradations.
This stabilization improves the temporal consistency of visual feedback available at the remote controller, which is essential for reliable closed-loop teleoperation. 

These observations indicate that mitigating ping-pong handovers by tuning mobility parameters can directly enhance the robustness of UE-to-controller video streaming.
Importantly, the performance improvement is achieved at the Non-RT timescale through explainable reasoning over RAN events and human-approved configuration updates, rather than through real-time actuation or black-box learning.

\section{Conclusion}
\label{sec:conclusion}

This paper
~\footnote{This research is partially carried out under the UK-Singapore BEACON project, funded by the UK Department for Science, Innovation and Technology (DSIT); and partially supported by the National Research Foundation, Singapore and Infocomm Media Development Authority under its Future Communications Research \& Development Programme. Any opinions, findings, and conclusions or recommendations expressed in this material are those of the author(s) and do not reflect the views of the funding agencies.} 
presents an LLM-based Net Analyzer rApp for the O-RAN Non-RT RIC that enables explainable, safe, and human-in-the-loop automation through event-informed, batch-triggered reasoning. The loop-driven architecture integrates RAN event analysis, operator interaction, and controlled decision-making, while the separation between reasoning and actuation—with mandatory operator approval—ensures bounded and transparent execution.
A real O-RAN testbed demonstrates the rApp’s end-to-end operation in a ping-pong handover scenario, showing how raw RAN telemetry is transformed into structured explanations and operator-approved parameter recommendations. Rather than replacing existing analytics or control mechanisms, the rApp augments them with an explanation-centric reasoning layer at Non-RT timescales that guides safe, human-supervised configuration updates.

We are going to extend the framework to include more use cases, incorporate richer cross-layer context, and evaluate scalability, highlighting the potential of LLMs as reliable reasoning co-pilots for 6G and beyond. %O-RAN management systems.

%\section*{Acknowledgment}
%\textbf{Acknowledgment:} This research is partially carried out under the UK-Singapore BEACON project, funded by the UK Department for Science, Innovation and Technology (DSIT); and partially supported by the National Research Foundation, Singapore and Infocomm Media Development Authority under its Future Communications Research \& Development Programme. Any opinions, findings, and conclusions or recommendations expressed in this material are those of the author(s) and do not reflect the views of the funding agencies.

\bibliographystyle{IEEEtran}
\bibliography{ref}
%\printbibliography
\end{document}